\documentclass[10pt,journal,compsoc]{IEEEtran}
\usepackage{subfigure}
\usepackage{graphicx}
\usepackage{subfig}
\usepackage[switch]{lineno}
\usepackage{amsmath}
\usepackage{amssymb}
\usepackage{multirow}
\usepackage{booktabs}

\usepackage{algorithm}
\usepackage{algorithmic}
\usepackage{xcolor}
\usepackage{xspace}
\usepackage{flushend}
\newcommand{\paratitle}[1]{\vspace{1.5ex}\noindent\textbf{#1}}
\newcommand{\ie}{\emph{i.e.,}\xspace}
\newcommand{\aka}{\emph{aka}\xspace}
\newcommand{\eg}{\emph{e.g.,}\xspace}
\newcommand{\baby}{{MMSBR}\xspace}
\newcommand{\babyx}{{MMSBR}}

\newcommand{\fig}{Fig.\xspace}

\ifCLASSOPTIONcompsoc
  \usepackage[nocompress]{cite}
\else
  \usepackage{cite}
\fi

\ifCLASSINFOpdf
\else
\fi

\begin{document}
%
\title{Beyond Co-occurrence: Multi-modal Session-based Recommendation}

\author{
Xiaokun Zhang, Bo Xu, Fenglong Ma, Chenliang Li, Liang Yang and Hongfei Lin
\IEEEcompsocitemizethanks{
\IEEEcompsocthanksitem Xiaokun Zhang, Bo Xu, Liang Yang and Hongfei Lin are with the School of Computer Science and Technology, Dalian University of Technology, Dalian, China.
E-mail: dawnkun1993@gmail.com, \{xubo, liang, hflin\}@dlut.edu.cn
\IEEEcompsocthanksitem Fenglong Ma is with the College of Information Sciences and Technology, Pennsylvania State University, Pennsylvania, USA. E-mail: fenglong@psu.edu
\IEEEcompsocthanksitem Chenliang Li is with the School of Cyber Science and Engineering, Wuhan University, Wuhan, China. E-mail: cllee@whu.edu.cn
}
\thanks{Manuscript received 16 Apr. 2023; revised 5 July 2023; accepted 26 Aug 2023. \\ 
(Corresponding author: Hongfei Lin.)}
}

%
%

\markboth{IEEE TRANSACTIONS ON KNOWLEDGE AND DATA ENGINEERING}
{ZHANG \MakeLowercase{\textit{et al.}}: }
%



\IEEEtitleabstractindextext{%

\begin{abstract}
Session-based recommendation is devoted to characterizing preferences of anonymous users based on short sessions. Existing methods mostly focus on mining limited item co-occurrence patterns exposed by item ID within sessions, while ignoring what attracts users to engage with certain items is rich multi-modal information displayed on pages. Generally, the multi-modal information can be classified into two categories: descriptive information (\eg item images and description text) and numerical information (\eg price). In this paper, we aim to improve session-based recommendation by modeling the above multi-modal information holistically. There are mainly three issues to reveal user intent from multi-modal information: (1) How to extract relevant semantics from heterogeneous descriptive information with different noise? (2) How to fuse these heterogeneous descriptive information to comprehensively infer user interests? (3) How to handle probabilistic influence of numerical information on user behaviors? 
To solve above issues, we propose a novel multi-modal session-based recommendation (\baby) that models both descriptive and numerical information under a unified framework. Specifically, a pseudo-modality contrastive learning is devised to enhance the representation learning of descriptive information. Afterwards, a hierarchical pivot transformer is presented to fuse heterogeneous descriptive information. Moreover, we represent numerical information with Gaussian distribution and design a Wasserstein self-attention to handle the probabilistic influence mode. Extensive experiments on three real-world datasets demonstrate the effectiveness of the proposed \baby. Further analysis also proves that our \baby can alleviate the cold-start problem in SBR effectively.

\end{abstract}

}

\maketitle

\IEEEdisplaynontitleabstractindextext

%
\IEEEpeerreviewmaketitle

\IEEEraisesectionheading{\section{Introduction}\label{sec:introduction}}

%
%
%
%

\IEEEPARstart{A}{s} an important tool to combat information overload, recommender system (RS) plays a vital role in present information era. Especially in context of e-commerce, RS facilitates online consumption by offering personalized services to individuals. Assuming the user identity information is accessible, conventional RS~\cite{IKNN,Koren@Computer2009} relies on user profiles and long-term behaviors to predict their preferences. However, in most real-world scenarios, the user identification is not available due to privacy policy or unlogged-in cases, where what RS could use is the short behavior sequences of anonymous users (\ie sessions). Apparently, conventional RS methods are no longer applicable or satisfactory in this case. To handle this situation, session-based recommendation (SBR) is proposed to predict next items interested by anonymous users within short sessions~\cite{Wang@CS2022}. Nowadays, SBR has drawn significant attention from both academia and industry due to its highly practical value~\cite{NARM,Han@SIGIR2022}.

With capacity in capturing transition patterns among items in a session, various neural networks are employed to improve SBR, such as recurrent neural networks (RNN)~\cite{GRU4Rec, NARM}, convolutional neural networks (CNN)~\cite{Yuan@WSDM2019}, attention mechanisms~\cite{STAMP, BERT4Rec}, and graph neural networks (GNN)~\cite{SR-GNN, LESSR}. Despite having made impressive progress, most existing methods still rely on mining \emph{co-occurrence patterns} exposed by item ID within short sessions. This significantly limits their performance since that a session usually contains a few items under the scenario of SBR (as shown in Table~\ref{statistics}, the average session length is no more than 3). In other words, there are not enough co-occurrence patterns for them to exploit for user intent modeling in SBR. Fortunately, the available \emph{multi-modal information} of items provides a promising antidote to improve SBR.

\begin{figure}[t]
  \centering
  \includegraphics[width=0.90\linewidth]{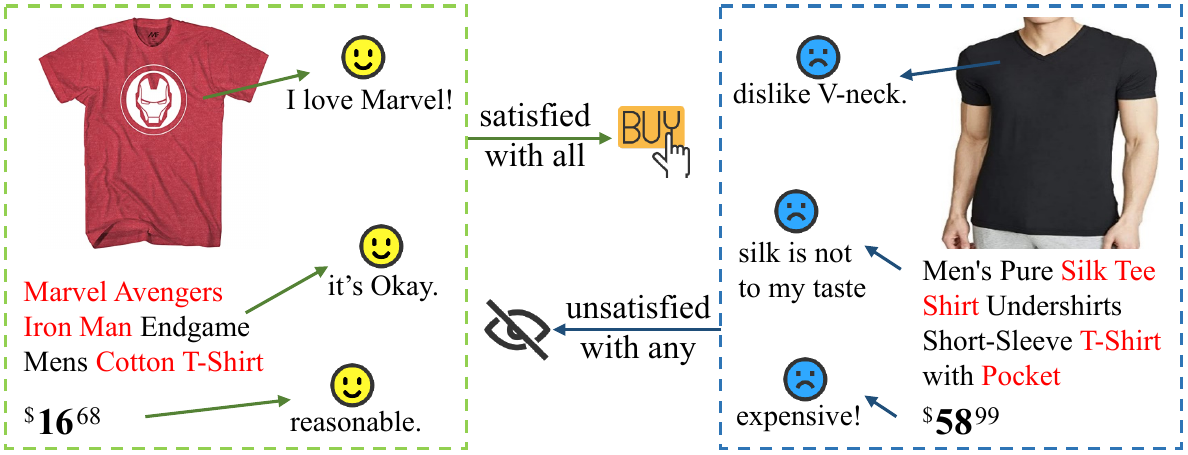}
  \caption{A user makes the decision after evaluating all multi-modal information displayed on pages including item images, description text and price. }\label{multimodal}
\end{figure}


Intuitively, it is the multi-modal information displayed on pages that drives users to engage with certain items. As shown in \fig~\ref{multimodal}, a user makes a decision usually after looking through item \emph{images}, reading description \emph{text}, and checking \emph{price}. 
Since relying on different vehicles for conveying particular item features, the multi-modal information can be categorized into two groups: 
\emph{descriptive} information and \emph{numerical} information. 
The descriptive information portrays an item with image and text that can intuitively describe some item features like style, color and material. For numerical information, \ie price, it delivers abstract value of an item through real numbers. In most cases, as illustrated in \fig~\ref{multimodal}, a user would not click an item unless she is satisfied with its all aspects. Obviously, the above multi-modal information jointly determines a user's choice.  

In fact, different from item ID that merely contains item co-occurrence patterns, multi-modal information presents extensive characteristics of items and encodes user fine-grained preferences. 
For example, a Marvel fan may have a high probability to purchase a T-shirt with the logo of iron man. Unfortunately, most existing models take neither images nor text into consideration, leading to their failure for accurate intent understanding. Moreover, co-occurrence based methods usually suffer from cold-start problem where there is no sufficient data to signify relations among new items~\cite{hou@KDD2022}. This issue will be smoothly solved, if we can understand user preferences from multi-modal features instead of dull item ID.
Although some recent models try to incorporate side information to facilitate user preferences learning, such as item category~\cite{lai@SIGIR2022}, description text~\cite{hou@KDD2022} and price~\cite{CoHHN}, they are still unable to reveal user intent holistically with such fragmentary information.
Thus, to fully understand user fine-grained preferences, we should consider the entire multi-modal information displayed on pages. However, it's nontrivial to utilize multi-modal information in SBR due to following obstacles:

(1) \emph{Descriptive information representation}. Under SBR scenario, images and text possess distinct noise. Normally, an item image not only contains the item for sale such as a cloth but also extra contents like accessories of the cloth.
Similarly, an item description text usually includes redundant words like exaggerated statements to attract user attention.
The existence of such noise in images and text increases the difficulty of extracting item semantics, hindering precise user preferences learning.
Therefore, the first challenge is how to obtain relevant semantics from heterogeneous descriptive information with different noise.

(2) \emph{Descriptive information fusion}. For an item, both image and text are utilized to describe its characteristics. Obviously, there exists shared information between them. At the same time, they also hold different purposes and focus on presenting distinct properties of items. To be specific, images are more intuitive than text to describe item colors and styles. Text can clearly express the material, \eg silk or cotton, whereas we can hardly understand it from images. Thus, the image and text complement each other and present an item in a united way. Accordingly, to comprehensively infer user interest, another challenge is how to fuse these heterogeneous descriptive information.


(3) \emph{Numerical information modeling}. In general, a user's taste is \emph{deterministic} on descriptive information. For instance, a user who prefers crewneck T-shirts may not click suggested ones with V-neck. In contrast, numerical price affects user behaviors in a \emph{probabilistic} way. More precisely, as long as the item price falls in a user's acceptable range, it does not matter if the price is slightly lower or higher. Thus, the last challenge is how to handle the probabilistic influence of numerical information on user behaviors.

In order to tackle above challenges, we propose a novel \underline{M}ulti-\underline{M}odal \underline{S}ession-\underline{B}ased \underline{R}ecommendation (\baby) that customizes both \emph{deterministic and probabilistic modelings} to handle descriptive and numerical information respectively. 
In the deterministic modeling, we devise a \emph{pseudo-modality contrastive learning} to refine descriptive information representations. In particular, contrastive learning is used to enhance representation learning by pushing semantically similar (positive) pairs close, while pulling dissimilar (negative) pairs apart~\cite{Wang@CIKM2022}. 
Since different modalities of an item refer to similar contents, it is intuitive to view them as positive pairs to tackle the noise issue. However, there are semantic gaps between distinct modalities, making it inappropriate to directly contrast them. To address this issue, we propose to utilize one modality to generate pseudo-information (namely pseudo-modality) in another modality via data generation techniques. The actual and pseudo modalities which are aligned in the same semantic space are then used as positive pairs in contrastive learning to mitigate noise existing in images and text. 

Moreover, to fuse descriptive information, we present a \emph{hierarchical pivot transformer} in deterministic modeling. With the ability in modeling complex relations in sequences, Transformer structure has shown to be effective for merging multi-modal signals~\cite{Attention, bottleneck}.
Inspired by this, we further create a pivot, which serves as a mixer of valuable information, in each transformer layer to govern the fusion of heterogeneous information. The pivot hierarchically extracts and integrates useful information from images and text under Transformer operations. We then view the pivot as the comprehensive embedding of descriptive information. 

In probabilistic modeling, we first represent item price as a \emph{Gaussian distribution embedding}, which enables \baby to perceive range property of item price. The \emph{Wasserstein self-attention} is then developed to handle price distribution embeddings for obtaining user acceptable price range. With the capacity in distinguishing differences between Gaussian distributions, the Wasserstein distance~\cite{ma@KDD2020,fan@WWW2022} is used in the Wasserstein self-attention to determine the relevance among price distribution embeddings. Finally, the proposed \baby provides personalized services for users via evaluating the entire multi-modal information displayed on pages.
In summary, the main contributions of our work are as follows:
\begin{itemize}
    \item We propose a novel \baby to characterize user preferences based on multi-modal information, which is more in line with user decision-making process than conventional co-occurrence based methods. To our best knowledge, this is the first work to reveal user intent from multi-modal information in SBR.
    \item We classify multi-modal information into descriptive and numerical types. Accordingly, we customize deterministic and probabilistic modeling that consist of several innovative techniques for comprehensively mining user intent.
    \item Extensive experiments over three public benchmarks demonstrate the superiority of \baby over state-of-the-art methods. Further analysis also justifies the effectiveness of \baby under cold-start scenario.
\end{itemize}


\section{Related Work}\label{sec:relate}
Considering that this work aims to improve session-based recommendation by incorporating multi-modal information, we briefly review the related work from following two aspects: (1) session-based recommendation including co-occurrence based methods and side information enhanced methods; (2) multi-modal recommendation.

\subsection{Session-based Recommendation}
\paratitle{Co-occurrence based methods}. 
Recent years, with the tremendous achievements of neural networks in various applications, we have witnessed the transition from traditional methods to neural models in SBR~\cite{Wang@CS2022}. With the intrinsic ability to handle sequential data, RNN as well as its variants are the first neural networks applied in SBR. For example, GRU4Rec~\cite{GRU4Rec} utilizes gated recurrent unit (GRU) to capture sequential patterns within sessions. NARM~\cite{NARM} enhances GRU4Rec with attention mechanism to explore user main purpose. Afterwards, many neural architectures are employed to model user sequential behaviors such as CNN~\cite{Yuan@WSDM2019}, attention mechanism~\cite{STAMP,BERT4Rec, DPAN4Rec}, GNN~\cite{SR-GNN,LESSR}, and reinforcement learning~\cite{Guo@TKDE2022}. Some approaches further enhance the learning for co-occurrence patterns via exploring extra sessions~\cite{CSRM,DIDN,Guo@KDD2019}, multi behaviors~\cite{yuan@ICDE2022}, multi user intents~\cite{Zhang@WSDM2023} and multi relations among items~\cite{WW@TKDE2023,Han@SIGIR2022}. Contrastive learning is an emerging technique whose target is to improve embeddings by enlarging the distance between positive and negative pairs. Many recent models utilize the technique to enable robust representation learning for accurate user intent modeling~\cite{Wang@CIKM2022,LXW@WSDM2023}. Also, other methods design new distance functions with metric learning to optimize user preferences learning~\cite{fan@WWW2022}. Although greatly promoting the development of SBR, all of these methods, essentially, focus on mining co-occurrence patterns reflected by item ID. They fail to perceive user fine-grained preferences concealing in multi-modal information, which becomes a bottleneck limiting their performance. 

\paratitle{Side information enhanced methods}. 
Considering that side information can help to unveil user unique taste, some methods try to incorporate various kinds of information to improve recommendation performance such as time (\aka positions)~\cite{Ye@SIGIR2020,wu@WWW2020}, categories~\cite{xie@SIGIR2022,lai@SIGIR2022}, price~\cite{CoHHN}, text~\cite{hou@KDD2022,liu@WWW2020} and images~\cite{Rashed@RecSys2022}. There are also a few works~\cite{MV-RNN, Pan@CIKM2022, Moreira@corr2021} taking both text and images into account to handle long sequential behaviors of users. These methods have proved the effectiveness of side information in understanding user interest. However, most of them conduct information fusion with simple concatenation or addition, leading to their failure in effectively merging various information. Moreover, they can neither alleviate noise in various modalities nor distinguish influence modes of distinct modalities on user behaviors. In addition, to our best knowledge, none of existing methods collectively considers entire multi-modal information displayed on the websites, \ie images, text and price, to simulate user behaviors. Thus, we propose a novel \baby to holistically reveal user intent from these multi-modal information, which is consistent with genuine decision-making process.

\subsection{Multi-modal Recommendation}
Multi-modal recommendation has received increasing attention recently, since that we humans perceive the world by concurrently processing and fusing multi-modal information~\cite{Tadas@TPAMI2019, bottleneck}. To name a few, some methods~\cite{Wei@MM2019,Wang@TM2023} employ Graph Neural Networks (GNN) and incorporate item images and text into the user-item interaction graph to facilitate user preferences and item characteristics learning. Beside, another line of research ~\cite{Liu@MM2021,Han@WWW2023} utilizes the pre-training technique to inject rich knowledge from item visual and textual modalities into recommender systems. More recently, BM3~\cite{Zhou@WWW2023} improves user and item representations by optimizing three multi-modal objectives including replicating user-item interaction graph and aligning modality features in inter- and intra-modality. Unfortunately, these methods fail to handle the situation of SBR because that they require users' identification and long-term behaviors to guide the model learning. Furthermore, there is no efforts bridging multi-modal information and SBR, hence we are the first to fill this research gap.


\section{Preliminaries}\label{sec:pre}
\begin{table}[t]
\centering
\caption{Important Notations.}
\renewcommand{\arraystretch}{1.2}
\begin{tabular}{ll}
\toprule
Notation      & Description \cr
\midrule
$\mathcal{I}$, $n$/$|\mathcal{I}|$       & item set, the total number of items  \cr
$x_i$ & an item      \cr
$\mathcal{S}=[x_1, x_2, ..., x_m]$    & a session with $m$ items \cr
$x_i^{img}$, $x_i^{txt}$, $x_i^{pri}$& item image, description text and price \cr
$v_i^{pri}$ & price encoding\cr
$\mathbf{e}_i^{img}$ & actual image embedding \cr
$\mathbf{e}_i^{txt}$ & actual text embedding \cr
$\mathbf{e}_{i}^{pseimg}$ & pseudo image embedding \cr
$\mathbf{e}_{i}^{psetxt}$ & pseudo text embedding \cr
$\mathbf{e}_i$ & descriptive information embedding \cr
$\mathbf{e}_{i}^{pri}$ & numerical information embedding \cr
$\mathbf{s}_d$ & representing user deterministic taste \cr
$\mathbf{s}_p$ & representing user acceptable price range

\\
                
\bottomrule
\end{tabular}
\label{notation}
\end{table}

\subsection{Problem Statement}
Session-based recommendation (SBR) is proposed to provide personalized services for anonymous users based on their short behavior sequences. Let $\mathcal{I}$ signify the set of all unique items, where $|\mathcal{I}| = n$ is the total number of items. Normally, as depicted in \fig~\ref{multimodal}, an item $x_i \in \mathcal{I}$ ($1 \leqslant i \leqslant n $) is presented to users in the form of multi-modal information including item image ($x_i^{img}$), description text ($x_i^{txt}$) and price ($x_i^{pri}$), \ie $x_i$ = \{$x_i^{img}$, $x_i^{txt}$, $x_i^{pri}$\}. In SBR, an anonymous user has chronologically interacted with $m$ items in a certain interval, producing a session $\mathcal{S}$ = [$x_1, x_2, ..., x_m$], where $x_m$ $\in$ $\mathcal{I}$. Our goal is to predict next item the user will prefer based on $\mathcal{S}$. Note that, we rely on rich multi-modal information users can access instead of dull item ID to reveal user intent, which enables our \baby to capture user fine-grained preferences and support cold-start scenario easily. The important notations used in this work are detailed in Table~\ref{notation}.

\subsection{Multi-modal Information Encoding}
Considering that distinct modalities are presented to users with completely different forms, \ie images in RGB, text in symbolic words and price in real numbers, we need to handle these information via special methods so that they can serve as inputs to neural models. In the next, we will detail how we encode these modalities, \ie image ($x_i^i$), text ($x_i^t$) and price ($x_i^p$).

\paratitle{Image embedding.} 
The first thing that a user may notice while browsing e-commerce websites is the item image. Due to the strong ability of GoogLeNet~\cite{googlenet} in extracting semantics from images~\cite{MV-RNN}, we apply it to obtain image embedding $\mathbf{e}_i^{img} \in \mathbb{R}^d$ from original image $x_i^{img}$ via,
\begin{align}
    \mathbf{e}_i^{img} &= {\rm imgEmb}(x_i^{img}),
\end{align}
where the ${\rm imgEmb}(\cdot)$ denotes the GoogLeNet model pre-trained on a large number of images. 

\paratitle{Text embedding.} 
After watching the image, the user further approaches the item by reading its description text. BERT~\cite{BERT} has been proved to be good at extracting text semantics by many studies~\cite{song@SIGIR2022,Liu@MM2021}. Therefore, we employ it to learn text embedding $\mathbf{e}_i^{txt} \in \mathbb{R}^d$ from original description text $x_i^{txt}$ via,
\begin{align}
    \mathbf{e}_i^{txt} &= {\rm textEmb}(x_i^{txt}),
\end{align}
where the ${\rm textEmb}(\cdot)$ denotes the BERT model pre-trained on large text corpus. 

\paratitle{Price encoding.} 
After evaluating descriptive image and text of an item, the user would check the item price to determine whether to purchase it. 
The absolute price cannot accurately indicate weather an item is expensive or not because that the price varies greatly across different categories (\eg tens of dollars for clothes and hundreds of dollars for electronics). Thus, for an item with price $x_i^{pri}$ in a certain category, we encode its price level via, 
\begin{align}
    v_i^{pri} = \lfloor \frac{x_i^{pri} - \min}{\max - \min} \times \rho \rfloor,
\end{align}
where [$\min, \max$] is the price range of its category, and $\rho$ is the total number of price levels. Notably, such a operation enables item price to be compared across different categories~\cite{CoHHN}.

\section{METHODOLOGY}\label{sec:method}

\begin{figure*}[t]
  \centering
  \includegraphics[width=0.9\linewidth]{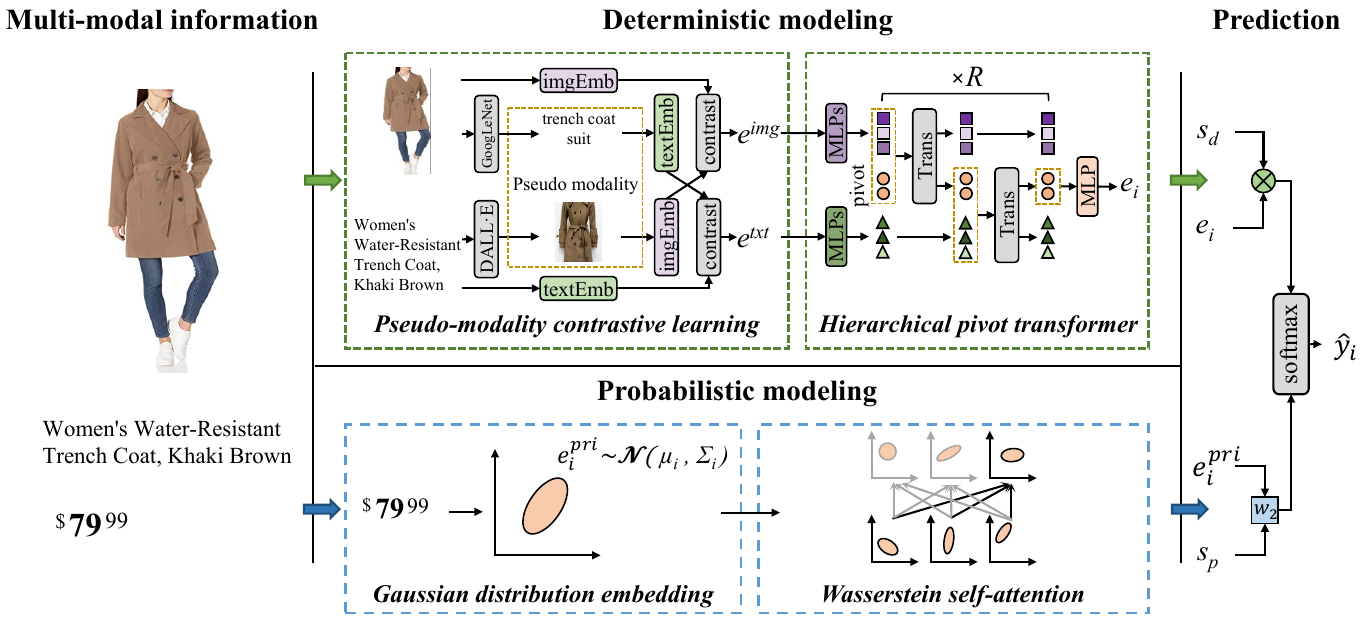}
  \caption{The proposed \baby customizes deterministic and probabilistic modeling to handle descriptive and numerical information respectively. In deterministic modeling, a pseudo-modality contrastive learning is designed to enhance descriptive information representations, a hierarchical pivot transformer is presented to fuse heterogeneous descriptive information, and a vanilla attention is used to capture user deterministic taste. The probabilistic modeling represents item price with Gaussian distribution embedding and devises Wasserstein self-attention to model user acceptable price range. Finally, we predict user behaviors based on the multi-modal information.}\label{MMSBR}
\end{figure*}

In this section, we will elaborate on the proposed \baby which is illustrated in \fig~\ref{MMSBR}. \baby is mainly composed of following interdependent components: (1) \emph{Deterministic modeling} is devised to handle descriptive information, \ie item image and description text, to capture user deterministic taste; (2) \emph{Probabilistic modeling} is developed to copy with numerical information, \ie item price, for modeling user acceptable price range; (3) \emph{Prediction} provides personalized services for individuals based on entire multi-modal information displayed on pages. 

\subsection{Deterministic Modeling}
Deterministic modeling employs: (1) \emph{pseudo-modality contrastive learning} to refine descriptive information representations; (2) \emph{hierarchical pivot transformer} to fuse heterogeneous descriptive information; (3) \emph{vanilla attention} to capture user deterministic taste.

\subsubsection{\textbf{Pseudo-modality Contrastive Learning}}
As stated before, there exists noise in item images and text, leading to inaccurate item semantics extraction. Contrastive learning can tackle this issue  by maximizing the agreements between semantically similar pairs. However, image and text embeddings coming from an item locate in distinct semantic spaces. Thus, it will corrupt the original semantics if we directly view them as positive pairs. To obtain effective contrastive signals, we resort to data generation techniques to generate pseudo modality which is aligned in the same space as the corresponding actual modality. Afterwards, with the generated contrastive signals, the contrastive learning is utilized to refine image and text embeddings. 

\paratitle{Pseudo-modality generation.} 
DALL·E~\cite{DALL} is an emerging technique to produce vivid images according to short text. For a piece of text $x_i^{txt}$, therefore, we feed it into the DALL·E to generate pseudo image $x_i^{pseimg}$. We then use the ${\rm imgEmb}(\cdot)$ to get the pseudo image embedding $\mathbf{e}_i^{pseimg} \in \mathbb{R}^d$ via,
\begin{align}
    \mathbf{e}_i^{pseimg} &= {\rm imgEmb}(x_i^{pseimg}).
\end{align}

As to the image $x_i^{img}$, we obtain its pseudo text by image classification. Concretely, we input $x_i^{img}$ into GoogLeNet to perform image classification with 1,000 categories, where each category label signifies a short text. The predicted top-$l$ categories, \ie a set of short texts, are then concatenated as pseudo text $x_i^{psetxt}$. Afterwards, we get the pseudo text embedding $\mathbf{e}_i^{psetxt} \in \mathbb{R}^d$ via,
\begin{align}
    \mathbf{e}_i^{psetxt} &= {\rm textEmb}(x_i^{psetxt}).
\end{align}


\paratitle{Contrastive learning.} 
The embeddings of actual and corresponding pseudo modalities, \ie $\mathbf{e}_i^{img}$ to $\mathbf{e}_i^{pseimg}$ (and $\mathbf{e}_i^{txt}$ to $\mathbf{e}_i^{psetxt}$), describe the same item and locate in the same semantic space. Naturally, we view them as positive pairs in contrastive learning to enhance image and text embeddings via, 
\begin{align}
\begin{split}
    \mathcal{L}_{con} &= -\frac{\exp({\rm sim}(\mathbf{e}_i^{img}, \mathbf{e}_i^{pseimg}))}{\sum_{k=1}^{n}\exp({\rm sim}(\mathbf{e}_i^{img}, \mathbf{e}_k^{pseimg}))} \\& - \frac{\exp({\rm sim}(\mathbf{e}_i^{txt}, \mathbf{e}_i^{psetxt}))}{\sum_{k=1}^{n}\exp({\rm sim}(\mathbf{e}_i^{txt}, \mathbf{e}_k^{psetxt}))},
\end{split}
\end{align}
where the ${\rm sim}(\cdot)$ is cosine similarity. In the first term, for an item image ($\mathbf{e}_i^{img}$), we view its pseudo image embedding ($\mathbf{e}_i^{pseimg}$) referring similar semantics as positives, while regarding other items' pseudo image embeddings ($\mathbf{e}_k^{pseimg}$) containing different contents as negatives. With pushing the positives close while pulling negatives apart, the \baby can enhance image embeddings. The second term does the same for refining text embeddings. With rich knowledge about corresponding modalities, the used data generation models not only align the positive pairs in the same space but also make pseudo modality contain core semantics of the actual modality. As shown in \fig~\ref{MMSBR}, the pseudo image retains core contents cloth and filters out redundant pants and shoes. Obviously, this is of benefit to the pseudo-modality contrastive learning for alleviating noisy information existing in distinct modalities.

\subsubsection{\textbf{Hierarchical Pivot Transformer}}
As demonstrated early, we need to fuse image and text features for comprehensive user interest understanding. Transformer structure has shown great potential in merging multi-modal signals, given that it can effectively mine complex relations among tokens in a sequence~\cite{Attention, bottleneck}. Inspired by this, we first apply several distinct MLPs to convert image/text embedding into different item feature embeddings and formulate feature sequence for image/text accordingly. Based on the feature sequences, a hierarchical pivot transformer is further proposed for effective descriptive information fusion.

\paratitle{Image/Text features generation.} 
We apply MLP to obtain feature embeddings because many studies have demonstrated the effectiveness of MLP in capturing semantics of input data~\cite{liu@ICMR2022,song@SIGIR2022}. Formally, an item image/text feature sequence ($\mathbf{Z}_{img}$/$\mathbf{Z}_{txt}$) is formulated via,
\begin{align}
    \mathbf{Z}_{img} &= \{\text{MLP}^{img}_1(\mathbf{e}_i^{img}),...,\text{MLP}^{img}_C(\mathbf{e}_i^{img})\}, \\
    \mathbf{Z}_{txt} &= \{\text{MLP}^{txt}_1(\mathbf{e}_i^{txt}),...,\text{MLP}^{txt}_C(\mathbf{e}_i^{txt})\}, 
\end{align}
where $\text{MLP}^{img}_k$ and $\text{MLP}^{txt}_k$ denote feed-forward neural networks with two hidden layers, and $C$ is the number of MLPs used for image/text features extracting. Note that, The $\text{MLP}^{img}_k(\mathbf{e}_i^{img})$ and $ \text{MLP}^{txt}_k(\mathbf{e}_i^{txt}) \in \mathbb{R}^{d}$ are certain feature embeddings of image and text respectively.

\paratitle{Hierarchical pivot transformer.} 
A vanilla transformer layer mainly contains three modules: Multi-head Self-Attention (MSA), Layer Normalisation (LN) and Fully Connected Layer (FCL). We can define a transformer layer with the input sequence $\mathbf{F}^l$ = [$f^{in}_1, f^{in}_2, ..., f^{in}_k$] and output sequence $\mathbf{F}^{l+1}$ =[$f^{out}_1, f^{out}_2, ..., f^{out}_k$] as $\mathbf{F}^{l+1} = \text{Trans}(\mathbf{F}^{l})$ via,
\begin{align}
    \mathbf{F}_*^l &= \text{MSA}(\text{LN}(\mathbf{F}^l)) + \mathbf{F}^l, \\
    \mathbf{F}^{l+1} &= \text{FCL}(\text{LN}(\mathbf{F}_*^l)) +\mathbf{F}_*^l.
\end{align}

Based on this, we further create a pivot $\mathbf{P}$ = [$\mathbf{p}_1,...,\mathbf{p}_T$] in each transformer layer to govern the fusion of multi-modal information, where $\mathbf{p}_i \in \mathbb{R}^{d}$ is a trainable token embedding used to assist information transmission. The hierarchical pivot transformer integrates the information of image ($\mathbf{Z}_{img}$) and text ($\mathbf{Z}_{txt}$) via:
\begin{align}
    [\mathbf{Z}_{img}^{l+1}, \mathbf{P}_{img}^l] &=\text{Trans}([\mathbf{Z}_{img}^l, \mathbf{P}^l]), \\
    \mathbf{p}_*^l &= (\mathbf{P}_{img}^l + \mathbf{P}^{l})/2, \\
    [\mathbf{Z}_{txt}^{l+1}, \mathbf{P}_{txt}^l] &=\text{Trans}([\mathbf{Z}_{txt}^l, \mathbf{P}_*^l]), \\
    \mathbf{p}^{l+1} &= (\mathbf{P}_{txt}^l + \mathbf{P}_*^l)/2,
\end{align}
where $\mathbf{P}^{0}$ = $\mathbf{P}$ (random initialization), $\mathbf{Z}_{img}^{0}$ = $\mathbf{Z}_{img}$ and $\mathbf{Z}_{txt}^{0}$ = $\mathbf{Z}_{txt}$. In each transformer layer, the pivot extracts and fuses important information from different modalities. Taking Eq.~(13) as an example, the pivot absorbs text information and transmits image information to the text modality. To fully fuse descriptive information, we stack the hierarchical pivot transformer defined by Eqs.~(11)-(14) $R$ times. Finally, the last layer pivot passed by a MLP is used to represent the descriptive information of an item $x_i$ as $\mathbf{e}_i \in \mathbb{R}^{d}$ via,
\begin{align}
    \mathbf{e}_i &= \text{MLP}(\mathbf{P}^R) = \text{MLP}([\mathbf{p}^R_1;\mathbf{p}^R_2; ...;\mathbf{p}^R_T]),
\end{align}
where $[;]$ denotes the concatenation operation, and $\text{MLP}$ is a feed-forward neural network with two hidden layers.

\subsubsection{\textbf{Vanilla Attention}} 
For an item $x_i$, we have obtained its embedding $\mathbf{e}_i$ for descriptive information involving image and text. Apparently, a user deterministic taste hidden in items she has interacted with. Thus, based on item sequence with descriptive information $\mathbf{E}_d$ = [$\mathbf{e}_1,\mathbf{e}_2,...,\mathbf{e}_m$], we can apply the vanilla attention as used in~\cite{SR-GNN,CoHHN} to obtain user deterministic taste $\mathbf{s}_d \in \mathbb{R}^{d}$ via,
\begin{align}
    \mathbf{s}_d &= \sum^{m}_{k=1} \alpha_k \mathbf{e}_k, \\
    \alpha_k &= \mathbf{u}\sigma(\mathbf{A}_1\mathbf{e}_k+\mathbf{A}_2\mathbf{\bar{e}}+\mathbf{b}),
\end{align}
where $\mathbf{A}_1$, $ \mathbf{A}_2  \in \mathbb{R}^{d \times d}$ and $\mathbf{b}$ are learnable parameters, $\mathbf{u}^T \in \mathbb{R}^d$ is a trainable vector used to determine items' importance in the session, and $\mathbf{\bar{e}}$=$\frac{1}{m}\sum_{k=1}^m \mathbf{e}_k$.

\subsection{Probabilistic Modeling} 
Probabilistic Modeling employs: (1) \emph{Gaussian distribution embedding} to represent item price; (2) \emph{Wasserstein self-attention} to model user acceptable price range.
\subsubsection{\textbf{Gaussian Distribution Embedding}}
As discussed before, user preferences on price present range instead of point-wise property. Therefore, we represent the price level $v_i^{pri}$ of an item $x_i$ with the Gaussian distribution via, 
\begin{align}
    \mathbf{\hat{e}}_{i}^{pri} &= \text{Gaussian}(v_{i}^{pri}) \sim \mathcal{N}(\hat{\mu}_{i}, \hat{\Sigma}_{i}),
\end{align}
where $\hat{\mu}_i$ and $\hat{\Sigma}_i \in \mathbb{R}^d$ are mean and covariance vectors respectively. \baby learns them with two distinct look-up embedding tables based on item price level. Note that, the mean and covariance vectors collectively signify the price range where the item falls in. 
As indicated in~\cite{CoHHN}, the user price preferences are related with item category, so we further incorporate category information to formulate price embedding $\mathbf{e}_{i}^{pri}$ for the item $x_i$ via, 
\begin{align}
    \mathbf{e}_{i}^{pri} \sim \mathcal{N}(\mu_{i}, \Sigma_{i}) &= \mathcal{N}(\hat{\mu}_{i} + \mathbf{e}^c_i, \hat{\Sigma}_{i} + \mathbf{e}^c_i),
\end{align}
where $\mathbf{e}^c_i \in \mathbb{R}^{d}$ is the category embedding of the item. It is noted that an item price is represented by Gaussian distribution instead of widely used point-wise vector embedding, which endows \baby with the ability to perceive range property of item price.

\subsubsection{\textbf{Wasserstein Self-attention}}

Self-attention is employed by various approaches~\cite{CoHHN, Attention} to model behavior sequences due to its capacity in capturing item-item transition patterns. However, the conventional self-attention calculates similarity between point-wise vector embeddings with dot product, which is unsuitable for our settings where the price is represented by Gaussian distribution. Therefore, we devise a Wasserstein self-attention which applies Wasserstein distance~\cite{ma@KDD2020,fan@WWW2022} to obtain attention scores between price distribution embeddings. Formally, the 
Wasserstein distance between two Gaussian distribution embeddings $\mathcal{G}_1 \sim \mathcal{N}(\mu_{1}, \Sigma_{1})$ and $\mathcal{G}_2 \sim \mathcal{N}(\mu_{2}, \Sigma_{2})$ is defined as, 
\begin{align}
    \mathcal{W}_2(\mathcal{G}_1, \mathcal{G}_2)=\sqrt{\left\|\mu_1-\mu_2\right\|_2^2+\left\|\left(\mathbf{\Sigma}_1\right)^{\frac{1}{2}}-\left(\mathbf{\Sigma}_2\right)^{\frac{1}{2}}\right\|_2^2}.
\end{align}

Referring to conventional self-attention, Wasserstein self-attention ($\text{WSA}$) handles price sequence $\mathbf{E}_p$=[$\mathbf{e}_1^{pri}$, $\mathbf{e}_2^{pri}$, ..., $\mathbf{e}_m^{pri}$] via,
\begin{align}
        \mathbf{H} &= \text{WSA}(A^Q\mathbf{E}_p, A^K\mathbf{E}_p, A^V\mathbf{E}_p), 
\end{align}
where $\mathbf{H}=\{\mathbf{h}_1, \mathbf{h}_2, ..., \mathbf{h}_m\} $ is output and $A^*$=($A^*_{\mu}, A^*_{\Sigma}$) ($*\in\{Q, K, V\}$) is used to map each distribution in $\mathbf{E}_p$ into query, key and value spaces respectively.  $A^*_{\mu}$ or $A^*_{\Sigma} \in \mathbb{R}^{d \times d}$ converts mean or covariance embeddings to corresponding spaces. Afterwards, the Wasserstein distance is employed to calculate the attention scores between query $A^Q \mathbf{e}_i^{pri}$ and key $A^K \mathbf{e}_j^{pri}$ via,
 \begin{equation}
    \begin{split}
    a_{ij} &= \mathcal{W}_2(A^Q \mathbf{e}_i^{pri}, A^K \mathbf{e}_j^{pri}) \\
    &= \mathcal{W}_2(\mathcal{N}(A^Q_{\mu}\mathbf{\mu}_i, A^Q_{\Sigma}\mathbf{\Sigma}_i), \mathcal{N}(A^K_{\mu}\mathbf{\mu}_j, A^K_{\Sigma}\mathbf{\Sigma}_j)).
    \end{split}
 \end{equation}

We then linearly sum up each value $A^V \mathbf{e}_j^{pri} \sim \mathcal{N}(A^V_{\mu}\mathbf{\mu}_j, A^V_{\Sigma}\mathbf{\Sigma}_j)$ according to its attention scores to $i$-th item price $a_{ij}$ to obtain the $i$-th output $\mathbf{h}_i \sim \mathcal{N}(\mathbf{h}_i^\mu, \mathbf{h}_i^\Sigma) $ via,
\begin{align}
    \mathbf{h}_i^\mu=\sum_{j=1}^m a_{i j} A^V_{\mu} \mathbf{\mu}_j, \text { and } \mathbf{h}_i^\Sigma=\sum_{j=1}^m a_{i j}^2 A^V_{\Sigma} \mathbf{\Sigma}_j.
\end{align}

Finally, the hidden state $\mathbf{h}_m$ is used to represent acceptable price range $\mathbf{s}_p$ for the user via,
\begin{align}
    \mathbf{s}_p = \mathbf{h}_m \sim \mathcal{N}(\mathbf{h}_m^\mu, \mathbf{h}_m^\Sigma).
\end{align}

\subsection{Prediction}
So far, for an item $x_i$, we have obtained its comprehensive representation ($\mathbf{e}_i, \mathbf{e}_i^{pri}$) based on its multi-modal information, where $\mathbf{e}_i$ is derived from descriptive information (image and text) and $\mathbf{e}_i^{pri} \sim \mathcal{N}(\mathbf{\mu}_i, \mathbf{\Sigma}_i)$ comes from numerical information (price). As to an anonymous user, $\mathbf{s}_d$ represents her deterministic taste on descriptive information, and $\mathbf{s}_p$ indicates her acceptable price range. Based on the entire multi-modal information displayed on
pages, thus, we can infer the probability of the user clicking item $x_i$ via,
\begin{align}
    \hat{y}_i &= softmax(\mathbf{e}_i \mathbf{s}_d + \mathcal{W}_{2}(\mathbf{e}_i^{pri}, \mathbf{s}_p)),
\end{align}
where we evaluate user deterministic behaviors with dot-product and user probabilistic behaviors with Wasserstein distance.
As in~\cite{NARM,SR-GNN,CoHHN}, we employ cross-entropy to improve recommendation performance via:
\begin{align}
    \mathcal{L}_{rec} = - \sum^n_{i=1} y_i \log (\hat{y}_i) + (1-y_i)\log(1-\hat{y}_i),
\end{align}
where $y_i$ is ground-truth label and $\hat{y}_i$ is predicted probability of item $x_i$ to be clicked. Finally, we train our \baby under the joint supervision of recommendation and contrastive learning via,
\begin{align}
    \mathcal{L} = \mathcal{L}_{rec} + \lambda \mathcal{L}_{con},
\end{align}
where the $\lambda$ is a constant controlling the strength of contrastive learning task.

\section{Experimental setup}\label{sec:setup}

\subsection{Research Questions}
We conduct extensive experiments to validate the effectiveness of \baby by answering following research questions:





\begin{itemize}
    \item \textbf{RQ1} Does the proposed \baby achieve state-of-the-art performance? (ref. Section~\ref{sec:overall})
    
    \item \textbf{RQ2} What is the effect of various novel techniques proposed in \baby? (ref. Section~\ref{sec:pseudo}-\ref{sec:probabilistic})
    
    \item \textbf{RQ3} What is the performance of \baby under cold-start scenario?  (ref. Section~\ref{sec:cold})
    
    
    \item \textbf{RQ4} How does session length influence the performance of SBR?  (ref. Section~\ref{sec:length})

    \item \textbf{RQ5} What is the influence of different modalities on the performance of SBR?  (ref. Section~\ref{sec:modality})
    
    \item \textbf{RQ6} What is the influence of key hyperparameters on \baby? (ref. Section~\ref{sec:hyperparameter})
    
\end{itemize}


\subsection{Datasets and Preprocessing}

\begin{table}[t]
\tabcolsep 0.2in 
\centering
\caption{Statistics of all datasets.}
\begin{tabular}{ccccc}
\toprule
Datasets      & Cellphones & Grocery & Sports \\
\midrule
\#item        & 8,614    & 11,638     & 18,796      \\
\#category    & 48      & 665        & 1,259       \\
\#interaction & 196,376   & 364,728  & 566,504     \\
\#session     & 78,026   & 127,548    & 211,959      \\
avg.length    & 2.52      & 2.86      & 2.67       \\
\bottomrule
\end{tabular}

\label{statistics}
\end{table}

\begin{table*}[ht]
\small
\tabcolsep 0.03in 
    \caption{Performance comparison of \baby with baselines over three datasets. The results (\%) produced by the best baseline and the best performer in each column are underlined and boldfaced respectively. Statistical significance of pairwise differences for \baby against the best baseline (*) is determined by the t-test ($p < 0.01$).}
    \renewcommand{\arraystretch}{1.1}
    \begin{tabular}{c cccc cccc cccc}  
    \toprule 
    \multirow{2}*{Method}& 
    \multicolumn{4}{c}{Cellphones}&\multicolumn{4}{c}{Grocery}&\multicolumn{4}{c}{Sports}\cr  
    \cmidrule(lr){2-5} \cmidrule(lr){6-9} \cmidrule(lr){10-13}
    &Prec@10&MRR@10&Prec@20&MRR@20 &Prec@10&MRR@10&Prec@20&MRR@20 &Prec@10&MRR@10&Prec@20&MRR@20\cr  
    \midrule  
    S-POP     &5.32&2.71&7.24&2.85	&20.65&17.00&23.64&17.25	&15.61&14.56&17.59&14.69\cr  
    SKNN    &21.07&9.95&24.71&10.21	&39.83&25.15&41.88&25.29	&31.79&21.31&33.86&21.46\cr  
    NARM    &20.59&15.32&24.12&15.56	&40.39&34.53&42.41&34.62	&31.64&26.94&34.17&27.12\cr  
    SASRec   &23.37&15.47&27.58&15.76	&40.97&34.76&43.02&34.92	&31.54&26.68&34.11&26.87\cr
    BERT4Rec   &22.28&14.39&27.09&14.73	&40.59&34.09&42.93&34.31	&31.57&26.85&34.32&27.07\cr
    SR-GNN      &21.80&15.60&25.08&15.77	&40.81&34.89&42.74&35.01	&31.96&27.43&34.29&27.51\cr  
    COTREC          &\underline{23.78}&10.82&\underline{28.33}&11.13    &41.28&30.60&43.24&30.75    &32.16&23.28&\underline{35.13}&23.46\cr
    MSGIFSR      &20.92&14.53&24.51&14.77	&41.34&35.25&43.40&35.47	&\underline{32.28}&\underline{27.56}&34.95&\underline{27.72}\cr 
     MGS           &21.74&14.29&25.21&14.54	&40.92&35.06&42.79&35.20	&31.63&26.75&33.76&26.89\cr 
     UniSRec    &22.73&15.36&26.65&15.63	&41.40&35.12&43.44&35.24	&31.90&26.91&34.41&27.04\cr
    CoHHN          &23.60&\underline{15.77}&27.71&\underline{15.96}	&\underline{41.58}&\underline{35.33}&\underline{43.59}&\underline{35.58}    &32.12&27.13&35.02&27.31\cr
    \midrule
    {\bf \baby}     &{ $ \bf 24.37^*$ }&{$\bf 16.47^*$}&{$\bf 29.22^*$ }&{$\bf 16.81^*$ }
	                &{ $ \bf 42.10^*$ }&{$\bf 35.91^*$}&{$\bf 44.27^*$ }&{$\bf 36.06^*$ }
	                &{ $ \bf 32.89^*$ }&{$\bf 28.10^*$}&{$\bf 35.64^*$ }&{$\bf 28.28^*$ }\cr
	\bottomrule
    \end{tabular}
    \label{performance}
    \vspace{5px}
\end{table*}  

We evaluate our \baby and all baselines on three datasets covering different characteristics and domains from Amazon\footnote{http://jmcauley.ucsd.edu/data/amazon/},  \ie Cell Phones and Accessories (\textbf{Cellphones}), Grocery and Gourmet Food (\textbf{Grocery}), as well as Sports and Outdoors (\textbf{Sports}). Following~\cite{CoHHN}, we organize user behaviors within one day to imitate SBR scenario. The last item in a session is taken as predicted target, and remaining items are used to model user intent. As in~\cite{NARM,SR-GNN}, we filter out sessions whose length is 1 and items appearing less than 5 times. Also, we delete items with missing or invalid images/text. We chronologically split each dataset into three parts with the ratio of 7:2:1 for training, validation and testing respectively. Relying on item ID, existing models~\cite{NARM,BERT4Rec,lai@SIGIR2022,Xia@CIKM2021,CoHHN} can not handle cold-start items which do not appear in training sets, so they simply delete these items from test sets. Following their settings, we also remove the cold-start items, where datasets' statistics is shown in Table~\ref{statistics}.  Besides, we retain cold-start items to investigate the performance of \baby under cold-start scenario in Section \ref{sec:cold}, where the cold-start situation is reported in Table~\ref{coldData}. 
Note that, although our setting is ubiquitous in real scenes, available datasets containing images, text and price are very scarce instead. Thus, we sincerely hope that our work can foster the development of multi-modal datasets for SBR.

\subsection{Evaluation Metrics}
As in~\cite{NARM,SR-GNN,CoHHN}, we evaluate the performance of \baby and baselines with following two widely used metrics: \textbf{Prec@k} (Precision) calculates the proportion of cases where the target item is within recommendation list; \textbf{MRR@k} (Mean Reciprocal Rank) is the average of reciprocal ranks of target item among recommendation list. Similar as ~\cite{Xia@CIKM2021,guo@WSDM2022,lai@SIGIR2022,CoHHN}, the k is set as 10 and 20 in this work. 

\subsection{Baselines}

The following two groups of competitive methods are selected as baselines for performance comparison:

\paratitle{Co-occurrence based methods} focus on mining item co-occurrence patterns to provide recommendation: 
\begin{itemize}
    \item \textbf{S-POP} recommends the most frequent items in the current session;

    \item \textbf{SKNN} predicts next items based on items' co-occurrence frequency in all sessions;

    \item \textbf{NARM}~\cite{NARM} utilizes GRU with attention mechanism to capture user main intent;

    \item \textbf{SASRec}~\cite{SASRec} applies Transformer architecture to model transitions among items;

    \item \textbf{BERT4Rec}~\cite{BERT4Rec} employs bidirectional self-attention to model user behaviors;

    \item \textbf{SR-GNN}~\cite{SR-GNN} captures complex relations among items via GNN;

    \item \textbf{COTREC}~\cite{Xia@CIKM2021} enhances item embeddings by contrastive learning.

    \item \textbf{MSGIFSR}~\cite{guo@WSDM2022} studies fine-grained co-occurrence relations by dividing a session into multiple snippets.
\end{itemize}

\paratitle{Side information enhanced methods} utilize extra information to facilitate user preferences learning: 
\begin{itemize}
    \item \textbf{MGS}~\cite{lai@SIGIR2022} exploits item categories for more accurate preferences estimation;

    \item \textbf{UniSRec}~\cite{hou@KDD2022} incorporates description text of items to obtain universal sequence representations;

    \item \textbf{CoHHN}~\cite{CoHHN} emphasizes the significance of price in determining user choices.
\end{itemize}

We have not included MML~\cite{Pan@CIKM2022}, which focuses on text and image-based long sequence learning, in our baselines. This decision was made because MML randomly deletes some items within a sequence during model training, which is unsuitable for SBR where a session typically consists of only a few items (as shown in Table~\ref{statistics}).



\subsection{Implementation Details}
To ensure fair comparison, we fix embedding size of all methods at 64. 
The other hyperparameters of \baby and all baselines are determined via grid search according to their performance on Prec@20 in validation set. For main hyperparameters of \baby, we investigate the number of stacked layers for hierarchical pivot transformer $R$ in $\{1, 2, 3, 4, 5\}$, the number of generated features $C$ in $\{2, 4, 6, 8, 10\}$ and the number of tokens in pivot $T$ in $\{1, 2, 3, 4, 5, 6\}$. 
Besides, we fix balance coefficient $\lambda$ = 0.01, retain top-2 ($l$=2) categories from image classification as pseudo text, and set the number of price levels as 100 ($\rho$ = 100) for all datasets. The mini-batch size and initial learning rate is $100$ and $0.001$, respectively. Given that the output dimension of GoogLeNet and BERT are 1024 and 768 respectively, we utilize PCA algorithm to reduce them to 64. 
We have released the source code of our work\footnote{{https://github.com/Zhang-xiaokun/MMSBR}}. 


\section{Results and Analysis}\label{sec:experiments}

\subsection{Overall Performance (RQ1)}\label{sec:overall}

We report the performance of \baby and all baselines in Table~\ref{performance}, where the following observations are noted:

(1) Among co-occurrence based methods, COTREC and MSGIFSR achieve competitive performance. We speculate that COTREC's good performance comes from its utilization of contrastive learning to improve session embeddings. As to MSGIFSR, it divides a session into many snippets containing consecutive items, enabling it to capture fine-grained co-occurrence relations among items.  

(2) For methods with side information enhancement, CoHHN (price) and UniSRec (text) have obvious advantages over MGS (category). As opposed to category, price and text are what users can immediately observe on item pages. This observation supports our claim that modeling what displays on websites is of benefit to capturing user intent. 

(3) Compared with co-occurrence based methods, the side information enhanced methods generally perform better. This signifies the validity of extra information in modeling user behaviors. It makes sense since that side information enables models to mine various user preferences, leading to effective intent understanding.

(4) Different baselines have varying performance on various datasets. Taking CoHHN as an example, it achieves the best performance on Grocery among all baselines, while its results on Sports left some to be desired. These methods just focus on a part of information that users may access, either item ID, category, text or price. In fact, however, a user evaluates all available information before making decisions. Therefore, they are incapable of capturing user preferences holistically, which results in their inferiority in discerning user intent across various context (\ie datasets). 

(5) The proposed \baby consistently outperforms all baselines in terms of all evaluation metrics on all datasets, which demonstrates its effectiveness for SBR. In particular, \baby surpasses the best baselines in Prec@20 and MRR@20 by 3.14\% and 5.33\% on Cellphones, 1.56\% and 1.35\% on Grocery and 1.45\% and 2.02\% on Sports. 
Given that a user makes decisions by evaluating item images, text and price, the modeling for entire multi-modal information in \baby is in line with decision process, contributing to revealing her intent more effectively. Besides, with reference to Table~\ref{statistics} and Table~\ref{performance}, we find that \baby obtains largest improvements in Cellphones that contains least items among all datasets. We argue that the introduction of multi-modal information enriches data and enables \baby to understand user demands from multiple perspectives. Therefore, the proposed \baby achieves impressive performance under the condition of sparsity data. It also reminds researchers that the multi-modal information is an antidote to copy with sparsity issue.

\subsection{Effect of Pseudo-modality Contrastive Learning (RQ2)}\label{sec:pseudo} 

\begin{figure}[t]
  \centering
  \includegraphics[width=0.9\linewidth]{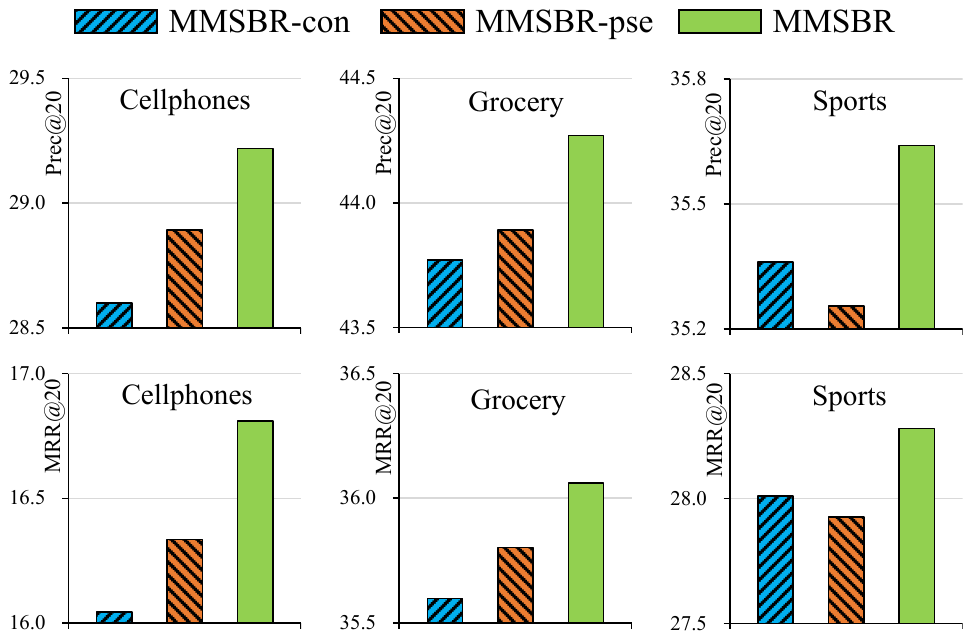}
  \caption{The effect of pseudo-modality contrastive learning.}\label{pseudo}
\end{figure}

To obtain relevant semantics from descriptive information under distinct noise, we propose a pseudo-modality contrastive learning. It refines image and text embeddings via contrastive learning, where generated pseudo modalities are used as contrastive signals. \babyx-con removes contrastive learning from \baby, \ie it directly fuses outputs of different modality encoders without handling distinct noise. \baby-pse projects embeddings of different modalities into a space via MLP and conducts contrastive learning accordingly like in~\cite{liu@ICMR2022,Wang@CIKM2022,LXW@WSDM2023}, while ignoring the semantic gaps existing in distinct modalities. 

As shown in \fig~\ref{pseudo}, in Cellphones and Grocery, both \baby and \baby-pse outperform \baby-con, demonstrating that contrastive learning can enhance modality representation. Besides, \baby-pse is defeated by \baby-con on Sports. It proves our hypothesis that semantic gaps between distinct modalities may impede representation learning. Thus, directly contrasting different modalities of an item in turn leads to performance degradation in this case. Moreover, \baby performs much better than \baby-pse in all datasets, which indicates that generated pseudo modalities can fill such semantic gaps. Additionally, \baby achieves the best performance across all variants, which indicates the superiority of pseudo-modality contrastive learning on mitigating distinct noise existing in different modalities. 


\subsection{Effect of Hierarchical Pivot Transformer (RQ2)}\label{sec:pivot} 

\begin{table}[t]
\tabcolsep 0.01in 
  \centering
    \caption{The effect of hierarchical pivot transformer.}
    \renewcommand{\arraystretch}{1.1}
    \begin{tabular}{c cc cc cc}  
    \toprule  
    \multirow{2}*{Method}& 
    \multicolumn{2}{c}{Cellphones}&\multicolumn{2}{c}{Grocery}&\multicolumn{2}{c}{Sports}\cr  
    \cmidrule(lr){2-3} \cmidrule(lr){4-5} \cmidrule(lr){6-7}
    &Prec@20&MRR@20 &Prec@20&MRR@20 &Prec@20&MRR@20\cr  
    \midrule  
    COTREC      &28.33&11.13    &43.24&30.75    &35.13&23.46\cr  
    MSGIFSR      &24.51&14.77	&43.40&35.47	&34.95&27.72\cr 
    \babyx$_{mlp}$     &26.74&15.95   &42.93&35.28    &34.67&27.86\cr
	{\bf \baby} &{ $ \bf 29.22^*$ }&{$\bf 16.81^*$} 
	            &{$\bf 44.27^*$ }&{ $ \bf 36.06^*$ }
	            &{$\bf 35.64^*$}&{$\bf 28.28^*$ }\cr
	\bottomrule
    \end{tabular}

    \label{pivot}
\end{table}

A user usually makes the decision after evaluating shared and distinct information from descriptive information. Therefore, we propose a novel hierarchical pivot transformer for heterogeneous information fusion. Following conventional operations~\cite{liu@ICMR2022,song@SIGIR2022}, \babyx$_{mlp}$ maps image and text into the same space by MLP and concatenates their embeddings to fuse item descriptive information. 

As shown in Table~\ref{pivot}, \babyx$_{mlp}$ is defeated by \baby with a large margin, which indicates the effectiveness of hierarchical pivot transformer in capturing complementary information from images and text. We believe that the pivot in each transformer layer is able to extract and integrate meaningful information from distinct modalities, thus facilitating effective information fusion. Furthermore, \babyx$_{mlp}$ achieves competitive performance (especially in MRR@20) compared with the best baselines MSGIFSR and COTREC. It serves as more evidence that modeling multi-modal information rather than only mining co-occurrence patterns of item ID can assist to better user intent learning.

\subsection{Effect of Probabilistic Modeling (RQ2)}\label{sec:probabilistic} 



\begin{figure}[t]
  \centering
  \includegraphics[width=0.9\linewidth]{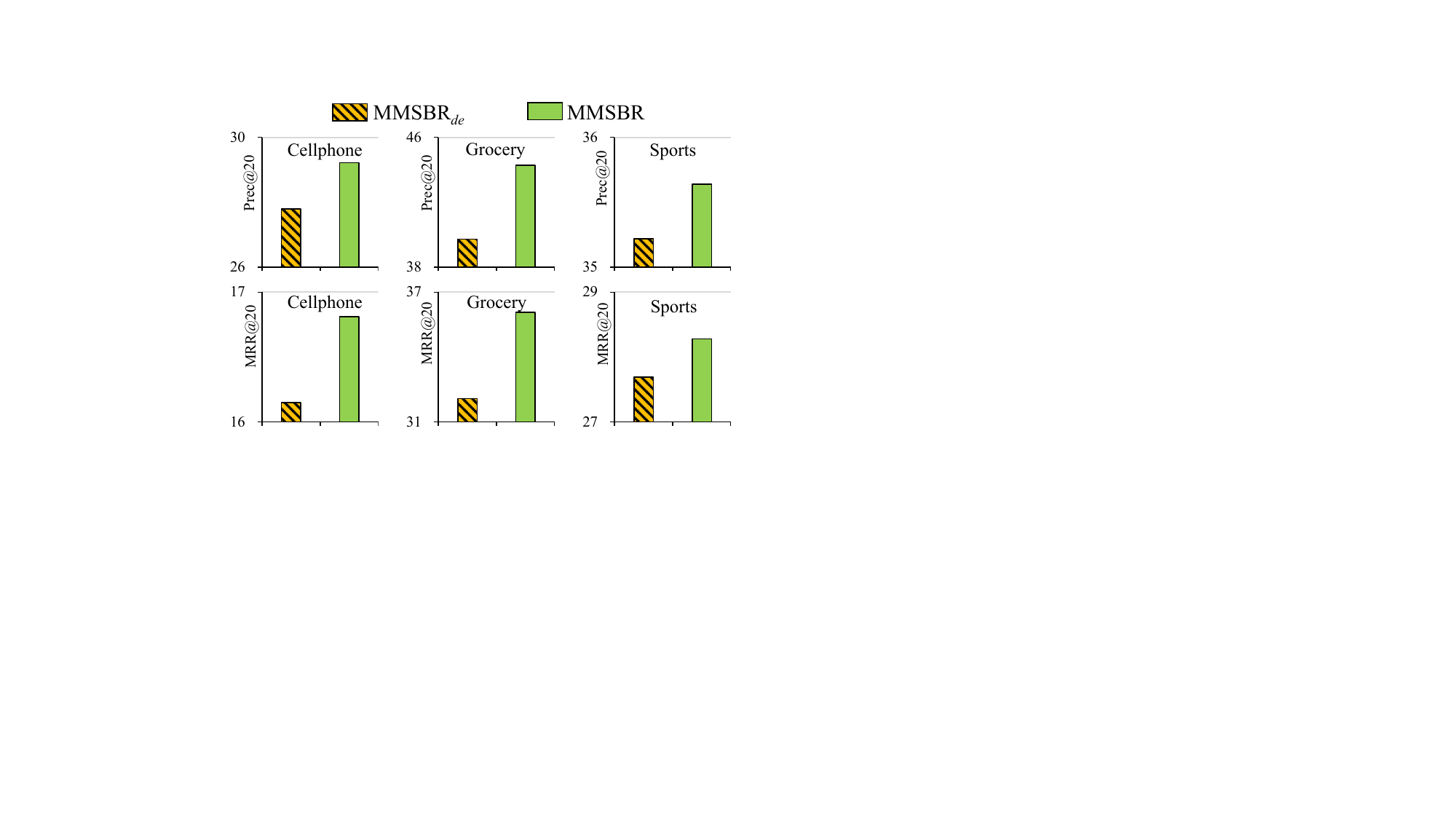}
  \caption{The effect of probabilistic modeling.}\label{probabilistic}
\end{figure}

As discussed previously, different from descriptive information where a user's taste is deterministic, the numerical information, \ie item price, affects user behaviors in a probabilistic way. Therefore, we propose a probabilistic modeling to handle this situation, where the Gaussian distribution and Wasserstein Self-attention are devised to represent item price and model user acceptable price range respectively. Following~\cite{CoHHN}, the variant \babyx$_{de}$ represents item price with point-wise vector embeddings instead of distribution ones. Specifically, \babyx$_{de}$ first discretizes continuous item price into discrete price-level, and then obtains point-wise embedding for the price via look-up embedding table. In other words, it does not discriminate distinct influence modes of various information on user choices. 

As presented in \fig~\ref{probabilistic}, \baby significantly outperforms \babyx$_{de}$ in all cases, confirming the validity of the proposed probabilistic modeling in tackling numerical information. Moreover, it demonstrates that users exhibit different behavioral patterns on different information, \ie deterministic/probabilistic mode on the descriptive/numerical information. By utilizing Gaussian distribution embeddings and Wasserstein self-attention, \baby is able to learn user acceptable price range, leading to its good performance on user behaviors modeling. In addition, distinguishing influence modes of different type information in a fine-grained manner is advantageous to user behaviors modeling, which is a valuable reference to future research.


\subsection{Performance in Cold-start Scenario (RQ3)} \label{sec:cold}

\begin{figure}[t]
  \centering
  \includegraphics[width=0.95\linewidth]{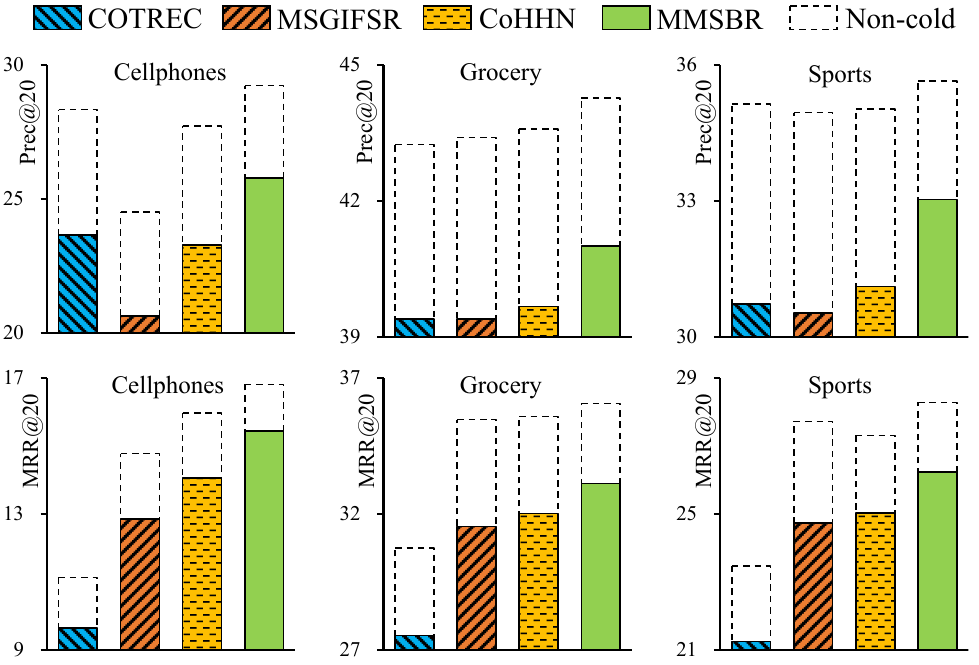}
  \caption{Performance in cold-start scenario.}\label{coldstart}
\end{figure}

Recommendation systems have long struggled with the cold-start problem, where they are required to show users new items that never appear in the system before.
To evaluate the performance of \baby under cold-start scenario, we retain fresh items which do not appear in training sets in tests, where the statistics is presented in Table~\ref{coldData}. We can get following insights from \fig~\ref{coldstart}, : 

\begin{table}[t]
\tabcolsep 0.05in 
\centering
\caption{Statistics of datasets with cold-start items.}
\renewcommand{\arraystretch}{1.1}
\begin{tabular}{ccccc}
\toprule
Datasets      & Cellphones+ & Grocery+ & Sports+ \\
\midrule
\#item        & 10,245(+1631)   & 13,493(+1855)    & 22,049(+3253)     \\
\#category    & 48(-)      & 678(+13)       & 1,312(+53)      \\
\#interaction & 199,065(+2689)  & 367,674(+2946) & 571,789(+5285)    \\
\#session     & 78,987(+961)  & 128,510(+962)   & 213,787(+1828)     \\
avg.length    & 2.52(-)     & 2.86(-)     & 2.67(-)      \\
\bottomrule
\end{tabular}

\label{coldData}
\end{table}

(1) When encountering with new items, all models show a deteriorated performance, indicating that the cold-start is truly a tricky issue in SBR. Fortunately, the incorporation of extra information can aid in portrayal for new items, leading to impressive performance in cold-start situation. For instance, in Sports, COTREC/MSGIFSR based on co-occurrence patterns defeats CoHHN incorporating price and category on Prec@20/MRR@20 for non-cold items, whereas CoHHN outperforms them both in cold-start scenario. 
(2) The co-occurrence based methods are prone to fail in cold-start scenario, which is intuitive as there are not co-occurrence patterns for them to learn. Solely relying on co-occurrence patterns exposed by item ID, these methods could do nothing but blindly guess user interest in new items with random embeddings, resulting in their inferior performance.
(3) The proposed \baby outperforms all methods with a large margin under cold-start scenario, indicating that \baby can effectively alleviate cold-start problem. Furthermore, our \baby has the least performance degradation compared with other methods in the cold start scenario. We believe that holistically modeling multi-modal information that a user can access enables \baby to mine her fine-grained preferences to the maximum, thus achieving impressive results. It also reminds researchers that utilizing multi-modal information is a promising way to copy with cold-start issue.

\subsection{Impact of Various Session Lengths (RQ4)}\label{sec:length} 

The session length can significantly affect recommendation performance since it signifies how much information we can obtain to model user intent.
Therefore, we investigate the performance of \baby under different session lengths. As shown in \fig~\ref{length}, following observations are noted: 
\begin{figure}[t]
  \centering
  \includegraphics[width=0.95\linewidth]{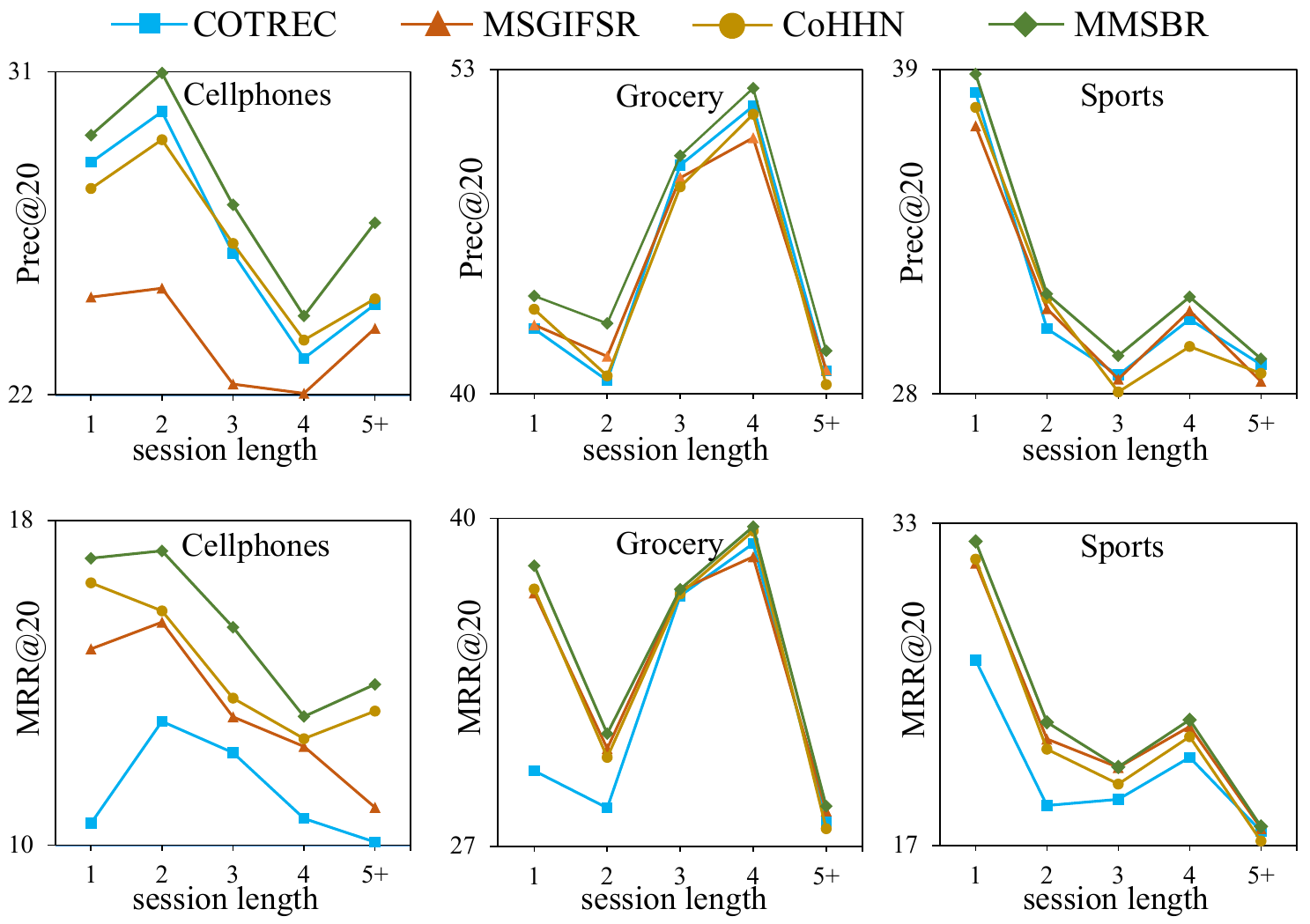}
  \caption{Impact of various session lengths.}\label{length}
\end{figure}

(1) The proposed \baby achieves larger improvement over baselines on short sessions ($\leqslant 3$) than long sessions ($>3$). Obviously, it is hard for co-occurrence based methods to accurately predict user behaviors within short sessions, since there is limited information for them to capture user intent. In contrast, our \baby can identify fine-grained preferences of users from rich multi-modal information, which alleviates data sparsity existing in short sessions. 
(2) Models perform better in short sessions than in long ones on Cellphones and Sports. Instead, they perform well in long sessions but poorly in short ones on Grocery. According to Table~\ref{statistics}, sessions in Grocery are much longer than that in Cellphones and Sports. We speculate that much more instances concentrated in long sessions make models achieve better performance in long sessions on Grocery.
(3) \baby achieves the best results in all cases, which demonstrates its effectiveness on modeling user behaviors in SBR again.


\subsection{Ablation Study (RQ5)}\label{sec:modality} 

\begin{table}[t]
\tabcolsep 0.01in 
  \centering
    \caption{The influence of different modalities.}
    \renewcommand{\arraystretch}{1.1}
    \begin{tabular}{c cc cc cc}  
    \toprule  
    \multirow{2}*{Method}& 
    \multicolumn{2}{c}{Cellphones}&\multicolumn{2}{c}{Grocery}&\multicolumn{2}{c}{Sports}\cr  
    \cmidrule(lr){2-3} \cmidrule(lr){4-5} \cmidrule(lr){6-7}
    &Prec@20&MRR@20 &Prec@20&MRR@20 &Prec@20&MRR@20\cr  
    \midrule  
    (a) w/o image      &27.45&14.85 &41.23&35.20    &32.14&27.50\cr  
    (b) w/o text    &27.19&14.69    &41.11&35.08    &32.22&27.42\cr 
    (c) w/o price    &25.10&13.35   &42.98&35.57    &34.78&27.68\cr
	{\bf \baby} &{ $ \bf 29.22^*$ }&{$\bf 16.81^*$} 
	            &{$\bf 44.27^*$ }&{ $ \bf 36.06^*$ }
	            &{$\bf 35.64^*$}&{$\bf 28.28^*$ }\cr
	\bottomrule
    \end{tabular}

    \label{modality}
\end{table} 

In this part, we further zoom into each modality to see its specific influence on \baby. We successively remove each modality from \baby to conduct ablation study. Notably, in (a)/(b) of Table~\ref{modality}, the item image/text is only used to refine text/image in pseudo-modality contrastive learning while we do not include it to model user interest.

As shown in Table~\ref{modality}, different modalities show various influence on \baby's performance in distinct context. For instance, without price, (c) is overwhelmed by (a) and (b) in Cellphones, while its performance is better than (a) and (b) in other datasets. We speculate that, for electronics, users are concerned with its price because there may be a huge price gaps between cellphones with different brands. As to Grocery, users tend to care its practicality instead of price. Moreover, \baby achieves much better performance than all variants. It supports our motivation that a user behaviors are determined by the entire multi-modal information displayed on pages. Thus, it is rationale and imperative to model user preferences by considering these multi-modal information holistically.

\subsection{Hyperparameter Study (RQ6)}\label{sec:hyperparameter}

In this section, we investigate the influence of three main hyperparamers on \baby. 

\begin{figure}[t]
  \centering
  \includegraphics[width=0.95\linewidth]{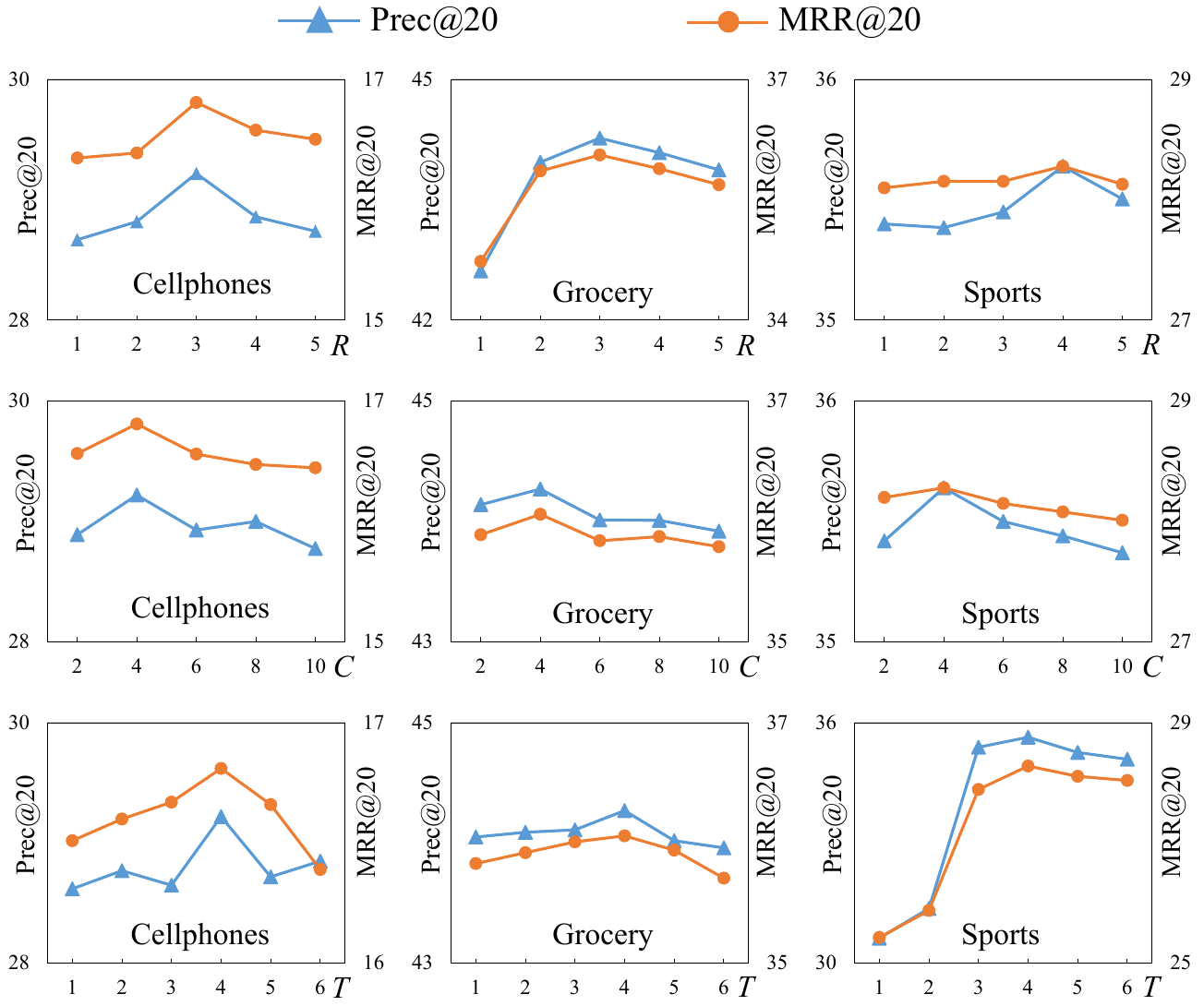}
  \caption{Impact of hyperparameters.}\label{hyperparameter}
\end{figure}

\paratitle{The number of stacked layers for hierarchical pivot transformer $R$.} From the first row in \fig~\ref{hyperparameter}, we can find that the optimal $R$ for Cellphones/Grocery and Sports is 3 and 4 respectively. As shown in Table~\ref{statistics}, Sports contains much more items than other datasets. We speculate that \baby needs to repeat hierarchical pivot transformer more times to fully integrate heterogeneous information in larger dataset. 

\paratitle{The number of generated features $C$.}
As shown in middle row in \fig~\ref{hyperparameter}, the small number of $C$, \ie 4, can make \baby achieve satisfactory results. It is consistent with cognitive anthropology that humans can only pay attention to a few aspects (features) of a matter (item) simultaneously. 

\paratitle{The number of tokens in pivot $T$.}
Refer to the last row in \fig~\ref{hyperparameter}, if $T$ is set too small, the pivot can not effectively extract information from different modalities. In contrast, if it is set too large, the information is sparsely distributed in each token, which is also adverse to information fusion. Accordingly, we empirically fix $T$ at 4 in all datasets.

\section{Conclusion and Future Work}\label{sec:conclusion}

Existing methods for session-based recommendation mostly concentrate on mining limited item co-occurrence patterns exposed by item ID, while ignoring that it is rich multi-modal information displayed on pages that attracts users to engage with certain items. Based on this motivation, we propose a novel \baby to characterize user preferences by modeling multi-modal information including descriptive information (images and text) and numerical information (price). Specifically, we devise a pseudo-modality contrastive learning to obtain relevant semantics of item images and text. Afterwards, a hierarchical pivot transformer is presented to effectively fuse heterogeneous descriptive information. For numerical information, we first represent item price with Gaussian distribution and devise a Wasserstein self-attention to model user acceptable price range. Comprehensive experiments conducted on three public datasets demonstrate the superiority of \baby over state-of-the-art baselines. Additional research also validates the effectiveness of \baby under cold-start scenario.

In the future, we plan to explore user reviews on items for further mining user fine-grained preferences for SBR. 
Besides, despite tailored for SBR, the proposed pseudo-modality contrastive learning and hierarchical pivot transformer can be easily extended to other multi-modal tasks for effective multi-modal learning.



\ifCLASSOPTIONcompsoc
  \section*{Acknowledgments}
\else
  \section*{Acknowledgment}
\fi
We thank the editors and reviewers for their efforts and constructive comments.
This work was supported by Natural Science Foundation of China (No.62076046, No.62006034).


\ifCLASSOPTIONcaptionsoff
  \newpage
\fi


\bibliographystyle{IEEEtran}
%

\bibliography{ref}



%
\vspace{-40px}
\begin{IEEEbiography}[{\includegraphics[width=1in,height=1.25in,clip,keepaspectratio]{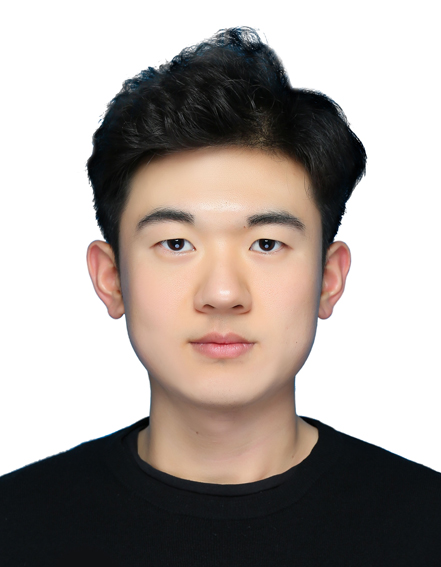}}]{Xiaokun Zhang}
is currently pursuing the PhD degree with the School of Computer Science and Technology, Dalian University of Technology, China. His research interests include data mining and information retrieval, mainly focusing on intelligent recommender systems.
\end{IEEEbiography}
\vspace{-40px}
\begin{IEEEbiography}[{\includegraphics[width=1in,height=1.25in,clip,keepaspectratio]{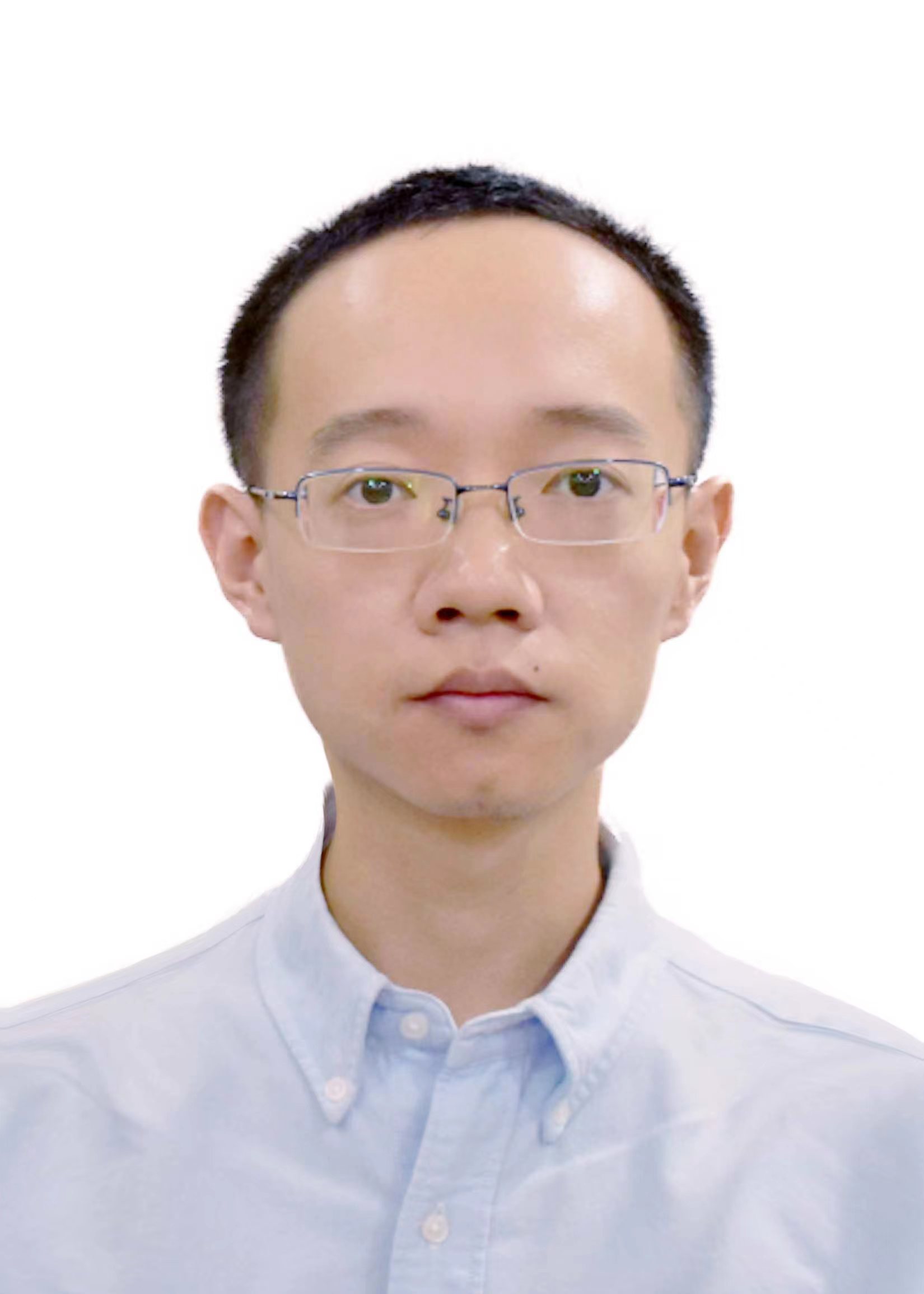}}]{Bo Xu}
received the BSc and PhD degrees from the Dalian University of Technology, China, in 2011 and 2018. He is currently an associate professor in School of Computer Science and Technology of Dalian University of Technology. His current research interests include information retrieval and natural language processing.
\end{IEEEbiography}
\vspace{-40px}
\begin{IEEEbiography}[{\includegraphics[width=1in,height=1.25in,clip,keepaspectratio]{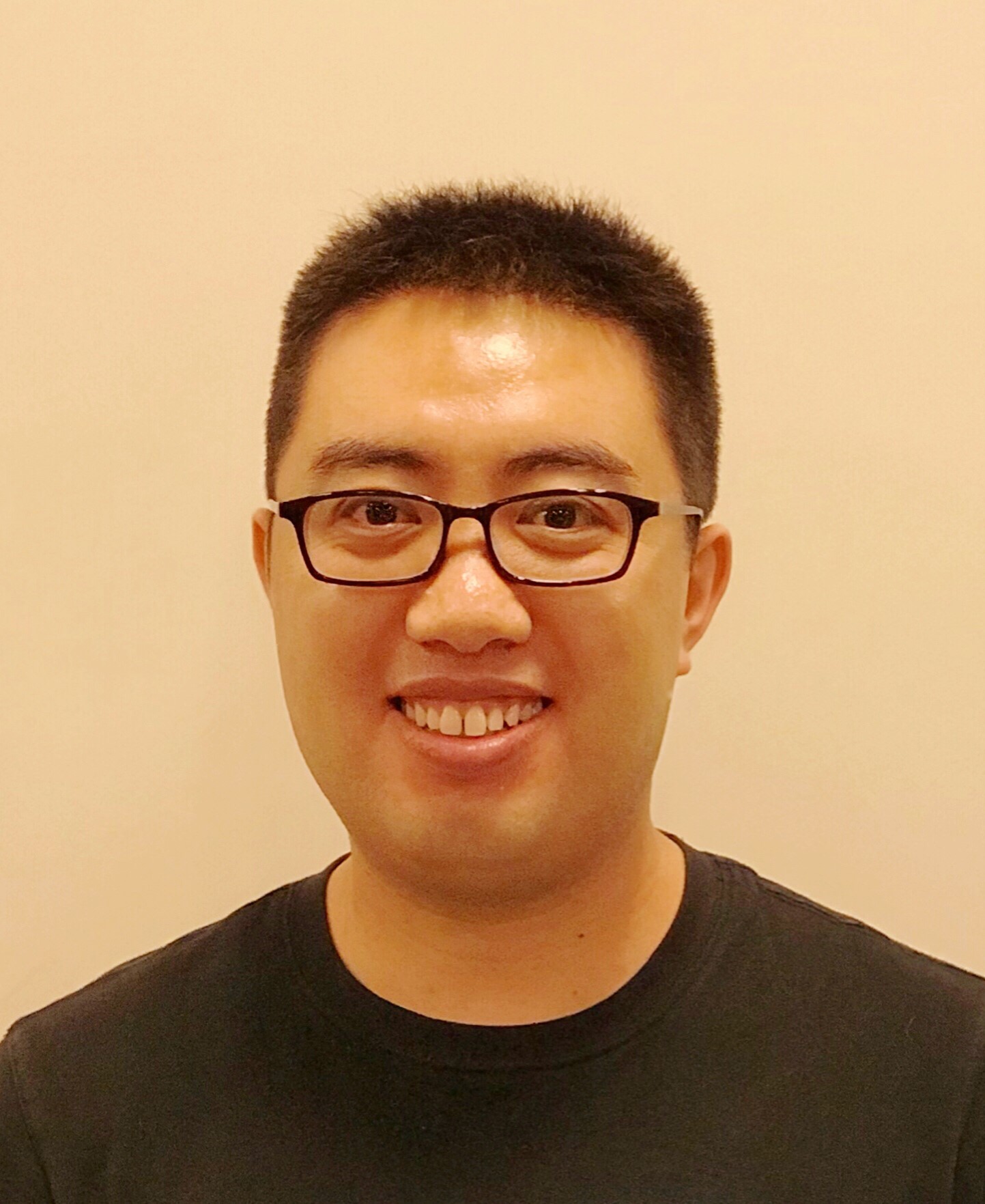}}]{Fenglong Ma}
is an assistant professor in the College of Information Sciences and Technology at the Pennsylvania State University. He received his Ph.D. from the Department of Computer Science and Engineering, University at Buffalo (UB) in 2019. His research interests lie in data mining and machine learning, with an emphasis on mining health-related data. His research interests also include natural language processing, social network mining and security.
\end{IEEEbiography}
\vspace{-40px}
\begin{IEEEbiography}[{\includegraphics[width=1in,height=1.25in,clip,keepaspectratio]{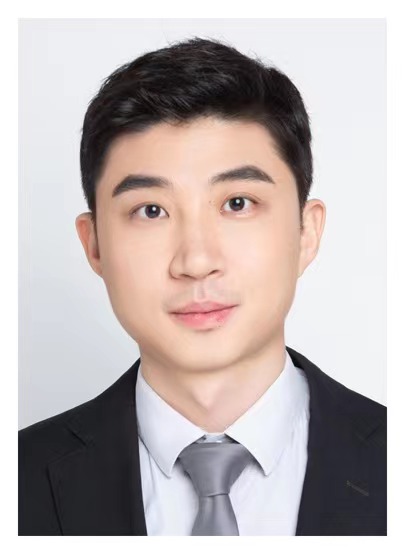}}]{Chenliang Li}
is a full Professor with School of Cyber Science and Engineering, Wuhan University China. His research areas include Information Retrieval, Recommendation
System, Natural Language Processing and Social Computing. He has published over 80 papers in leading conferences and journals, and was the recipient of SIGIR 2016 Best Student Paper Award Honorable Mention. Currently, He serves as an Associate Editor for ACM TOIS, ACM TALLIP, and an editorial board member of JASIST and IPM. He has been a PC member of many leading conference, such as SIGIR, WWW, ACL, WSDM, AAAI, CIKM.
\end{IEEEbiography}
\vspace{-40px}
\begin{IEEEbiography}[{\includegraphics[width=1in,height=1.25in,clip,keepaspectratio]{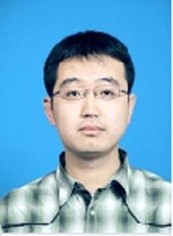}}]{Liang Yang}
received the BSc and PhD degrees from the Dalian University of Technology, China, in 2009 and 2017, respectively. He is currently a lecturer in School of computer science and technology at the Dalian University of Technology. His current research interests include sentiment analysis and text mining.
\end{IEEEbiography}
\vspace{-40px}
\begin{IEEEbiography}[{\includegraphics[width=1in,height=1.25in,clip,keepaspectratio]{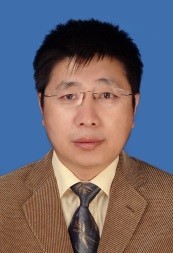}}]{Hongfei Lin}
received the BSc degree from the Northeastern Normal University in 1983, the MSc degree from the Dalian University of Technology in 1992, and the PhD degree from the Northeastern University in 2000. He is currently a professor in School of Computer Science and Technology at the Dalian University of Technology. He has published more than 500 research
papers in various journals, conferences, and books. His research interests include information retrieval, text mining for biomedical literatures, biomedical hypothesis generation, information extraction from huge biomedical resources, learning-to-rank. He is the director of Information Retrieval Lab. at Dalian University of Technology.
\end{IEEEbiography}



\end{document}